\documentstyle[preprint,aps]{revtex}
\begin{document}
\draft
\title{Neutrino-induced neutron spallation and supernova\\
$r$-process nucleosynthesis}
\author{Y.-Z. Qian$^1$, W. C. Haxton$^2$, 
K. Langanke$^{3,4}$, and P. Vogel$^1$}
\address{$^1$Department of Physics, 161-33,
	California Institute of Technology,\\
        Pasadena, California 91125}
\address{$^2$Institute for Nuclear Theory, Box 351550, 
and Department of Physics, Box 351560,\\
	University of Washington, Seattle, Washington 98195}
\address{$^3$W. K. Kellogg Radiation Laboratory, 106-38,
	California Institute of Technology,\\
        Pasadena, California 91125}
\address{$^4$Institute of Physics and Astronomy, 
         Univeristy of Aarhus, Denmark}

\date{\today}
\maketitle
\begin{abstract}
It is quite likely that the site of the $r$-process is the hot,
neutron-rich ``bubble" that expands off a protoneutron star during a
core-collapse supernova.  The $r$-process would then occur in an
intense flux of neutrinos.  In order to explore the consequences
of the neutrino irradiation,
we calculate the rates of charged-current and neutral-current 
neutrino
reactions on neutron-rich heavy nuclei, and estimate the
average number of neutrons emitted in the
resulting spallation.  Our results suggest, for a dynamic $r$-process
occurring in an expanding bubble, that charged-current $\nu_e$
captures might help shorten the timescale for the $r$-process, 
bringing it into better accord with our expectations about the
conditions in the hot bubble: neutrino reactions can be important in 
breaking through the waiting-point nuclei at $N=50$ and 82, while
still allowing the formation of abundance peaks.

Furthermore, after the $r$-process freezes out, there appear to 
be distinctive neutral-current and 
charged-current postprocessing effects.
These include a spreading of the abundance peaks and damping 
of the most pronounced features (e.g., peaks and valleys) in the
unpostprocessed abundance distribution.
    
Most importantly, a subtraction of the neutrino postprocessing 
effects from the observed solar $r$-process abundance
distribution shows that
two mass regions, $A=124$--126 and 183--187, are inordinately
sensitive to neutrino postprocessing effects.  This imposes
very stringent bounds on the freeze-out radii and dynamic
timescales governing the $r$-process.  Moreover, we find 
that the abundance patterns within these mass windows are entirely
consistent with synthesis by neutrino interactions.  This
provides a strong argument that the $r$-process must occur 
in the intense neutrino flux provided by a core-collapse
supernova.  It also greatly restricts dynamic models for
the supernova $r$-process nucleosynthesis.
\end{abstract}
\pagebreak

\section{Introduction}

It is known that approximately half 
of the heavy elements with $A>70$ and all
of the transuranics are formed by the process of rapid neutron
capture, the $r$-process.  The astrophysical site where the required
conditions for the $r$-process are produced --- neutron number
densities in excess of $\sim10^{20}\ {\rm cm}^{-3}$ and temperatures
of $\sim10^9$ K
lasting for on the order of 1 s \cite{mathews}
--- has been a matter of
continuing speculation.  Recently it
has been argued that an attractive and plausible site is the
``hot bubble'' that expands off the protoneutron star during a
core-collapse supernova \cite{woosley}.  
Neutron-rich matter initially composed
of free nucleons is blown off the neutron star.
As this nucleon soup expands and cools, almost all the protons are
locked into $\alpha$-particles.  Then an $\alpha$-process takes place
to burn $\alpha$-particles into seed nuclei 
with $A$ close to 100 \cite{woohoff}.
The $r$-process occurs through the capture of the excess neutrons
on these seed nuclei.  Although
numerical calculations of this process fail in some aspects,
both the produced $r$-process abundance 
distribution and the amount of 
$r$-process material ejected per supernova are roughly in accord
with observation \cite{woosley}.

If the $r$-process occurs near the protoneutron star, within 
perhaps 1000 km, then it takes place in an intense flux of 
neutrinos of all flavors emitted by the cooling neutron star.
In this paper we study the effects of charged-current and
neutral-current neutrino reactions with neutron-rich heavy nuclei
during and following the $r$-process.  Neutrino reactions can
affect the $r$-process nucleosynthesis
in two ways, by driving nuclear transitions
that alter the path or pace of the $r$-process, or by
modifying the abundance pattern through neutrino-induced neutron 
spallation after the $r$-process is completed.
During the $r$-process perhaps the most interesting possibility
is charged-current $\nu_e$ capture at a rate competitive with
$\beta$-decay, which would therefore speed up the nuclear
flow from one isotopic chain to the next of higher $Z$ \cite{nadyo}.  
This could be quite helpful in accelerating the passage through the
closed-neutron-shell nuclei at $N=50$ and 82, as conventional 
waiting times of several seconds are perhaps a bit troublesome
in relation to the shorter hydrodynamic timescales for the hot bubble.
(Note however, the nuclear flow through the $N=50$ closed neutron
shell may be carried by $\alpha$-capture reactions during the
$\alpha$-process preceding the $r$-process as in the particular
scenario of Woosley {\it et al.} \cite{woosley}.)  On the
other hand, it is clear that the existence of 
abundance peaks at $A\sim80$,
130, and 195 places some constraints on this possibility:
these peaks are clear signatures that slow 
waiting-point $\beta$-decay
rates are controlling the nuclear flow at the time the $r$-process
freezes out \cite{kratz}.  
By comparing the $\beta$-decay rates with the 
flux-dependent $\nu_e$ capture rates, Fuller and Meyer \cite{fuller} 
showed that the individual abundance peaks have to be made 
at sufficient distances above the neutron star.  In our study we
have extended their work and that of McLaughlin and Fuller \cite{gail}
by considering the competition between $\beta$-decay
and charged-current neutrino reactions in the context of a dynamic,
expanding bubble in order to more realistically determine what
role neutrinos may play in the nuclear flow.

However, our present work is mainly concerned with the postprocessing
of the $r$-process abundance distribution by neutrino-induced neutron
spallation.  Apart from the study of 
Domogatskii and Nad\"{e}zhin \cite{domog},
who estimated production yields for certain bypassed isotopes
from charged-current neutrino spallation reactions, little 
work has been done on this possibility.  We find that the 
spallation following charged-current 
and neutral-current neutrino excitation of 
nuclei can have a number of effects on $r$-process yields:
abundance peaks can be shifted and broadened, 
minima in the abundance pattern can be filled, and
the unevenness of yields can be smoothed in a characteristic way.
But most importantly, one can demonstrate directly from the 
neutrino physics calculations and the known $r$-process abundance
distribution that two mass windows, $A=124$--126 and 183--187,
are inordinately sensitive to neutrino postprocessing.  
This imposes new and stringent bounds on the freeze-out radii
and dynamic timescales in the supernova 
$r$-process models.  Moreover, the
pattern of abundances within these windows is entirely
consistent with neutrino-induced synthesis.  Unless this is
an unfortunate and accidental result, it would appear to provide 
direct proof that the $r$-process occurs in an intense 
neutrino fluence, i.e., in a core-collapse supernova.  
Further knowledge of
postprocessing neutrino fluences 
greatly reduces the freedom in dynamic
models of the $r$-process, relating, for example, the dynamic
timescale to the conditions at freeze-out.
These results take on added significance because the neutrino
physics of a core-collapse supernova is 
generally believed to be far less
model dependent than the hydrodynamics.

In this paper we also calculate the nuclear physics input
for the postprocessing and other possible effects of neutrinos
on the $r$-process. We present
detailed estimates of the charged-current $\nu_e$ capture rates 
using both empirical data and the shell model, the 
comparison of which provides some measure of the nuclear
structure uncertainties.  
The continuum random phase approximation (CRPA)
and shell-model techniques 
are employed in the calculations of neutral-current cross sections.
The subsequent decay of the highly-excited nuclei by neutron
emission is estimated by the statistical Hauser-Feshbach techniques.
We stress that the distinctiveness of the neutrino postprocessing
signals can be traced to well-understood aspects of nuclear
structure, such as the tendency of 
highly-excited neutron-rich nuclei to
emit multiple neutrons.

\section{Calculations of neutrino-nucleus interaction processes}

The rate for a specific neutrino reaction at a distance $r$ from
the center of the neutron star can be expressed
in convenient units as
\begin{equation}
\lambda_\nu\approx 4.97
\left({L_\nu\over 10^{51}\ {\rm erg\ s}^{-1}}\right)
\left({{\rm MeV}\over\langle E_\nu\rangle}\right)
\left({100\ {\rm km}\over r}\right)^2
\left({\langle\sigma_\nu\rangle\over 10^{-41}\ 
{\rm cm}^2}\right)\ {\rm s}^{-1},
\end{equation}
where $L_\nu$ and $\langle E_\nu\rangle$ are 
the luminosity and average energy, respectively, 
of the neutrino species responsible for the reaction, and
$\langle\sigma_\nu\rangle$ is the 
corresponding cross section averaged
over the neutrino spectrum.
The spectrum-averaged neutrino reaction cross
section is 
\begin{equation}
\langle\sigma_\nu\rangle = \sum_f\int \sigma_\nu^f(E_\nu)
f_\nu(E_\nu)dE_\nu,
\end{equation}
where the sum extends over all possible final nuclear states $f$.
The neutrino spectrum in the above equation,
$f_\nu(E_\nu)$, is taken to be
\begin{equation}
f_\nu(E_\nu) =  {1\over F_2(\eta_\nu)T_\nu^3}{E_\nu^2\over 
\exp[(E_\nu/T_\nu)-\eta_\nu]+1},
\end{equation}
where $T_\nu$ and $\eta_\nu$ are parameters fitted to numerical 
spectra, and $F_2(\eta_\nu)$ normalizes the spectrum to unit flux.
The transport calculations of Janka \cite{janka} yield spectra with
$\eta_\nu \sim 3$ for all neutrino species. 
While this choice also provides a good fit to the $\nu_e$ and
$\bar\nu_e$ spectra calculated by Wilson and Mayle, 
their heavy-flavor neutrino spectra
have approximately a black-body shape ($\eta_\nu \sim 0$)
\cite{wilson}.  In this work we take $\eta_{\nu_e}=3$ and 
$\eta_{\nu_{\mu(\tau)}} = \eta_{\bar\nu_{\mu(\tau)}}=0$, 
though we also have done
calculations with other choices of $\eta_\nu$.  
The average neutrino energy is given by
$\langle E_\nu\rangle\approx 3.99T_\nu$ for $\eta_\nu = 3$, and
$\langle E_\nu\rangle\approx 3.15T_\nu$ for $\eta_\nu = 0$.

The neutrino interactions of interest are charged-current $\nu_e$
and neutral-current $\nu_{\mu(\tau)}$ 
and $\bar\nu_{\mu(\tau)}$ reactions
on the waiting-point nuclei in the $r$-process. 
For the very neutron-rich
heavy nuclei in the $r$-process, 
charged-current $\bar\nu_e$ reactions
can be neglected because allowed transitions are Pauli blocked.
For typical average supernova neutrino energies
$\langle E_{\nu_e}\rangle\approx 11$ MeV, 
$\langle E_{\bar\nu_e}\rangle\approx 16$
MeV, and $\langle E_{\nu_{\mu(\tau)}}\rangle\approx 
\langle E_{\bar\nu_{\mu(\tau)}}\rangle\approx 25$ MeV,
the heavy-flavor neutrinos dominate 
neutral-current reaction rates. (However, for these average
energies, the charged-current $\nu_{\mu(\tau)}$ reactions
are energetically prohibited.)

Thus the task before us is to estimate the 
relevant cross sections for
an appropriate range of neutrino energies and final nuclear states,
so that Eq. (2) can be evaluated.  In principle this could be done
explicitly for each nucleus involved in the $r$-process network,
using some technique like the CRPA or the shell model.  But this 
approach is clearly impractical numerically.  Instead, we 
present in this section a more schematic description of the
cross sections, one largely based on experimental systematics.
Our strategy is then to check 
this phenomenological approach with explicit
CRPA and shell-model calculations for a few test nuclei.

At typical supernova neutrino energies 
the dominant contribution to the total cross section for
$\nu_e$ capture on a parent nucleus with charge $Z$
comes from the allowed transitions to the isobaric
analog state (IAS) and the Gamow-Teller (GT) resonance states in the
daughter nucleus.  The allowed cross section is 
\begin{equation}
\sigma(E_{\nu_e}) = 
{G^2_F \cos^2\theta_c \over \pi} k_e E_e F(Z+1,E_e) 
[|M_F|^2 + (g_A^{\rm eff})^2 |M_{GT}|^2],
\end{equation}
where $G_F$ is the Fermi constant, $E_e$ and $k_e$ the energy and
three-momentum of the outgoing electron, respectively,
$\theta_c$ the Cabibbo 
angle, and $F(Z+1,E_e)$ a correction for the Coulomb distortion
of the outgoing electron wavefunction. The relativistic
form of $F(Z,E_e)$ is used in evaluating the integral in
Eq. (2). We also use an effective
axial-vector coupling constant $g_A^{\rm eff}$ = 1, rather than the
bare nucleon value 1.26, a renormalization that improves the
agreement between shell-model studies and experiments in the
$2s1d$ and $2p1f$ shells \cite{brown}.   

The allowed Fermi and GT transition strengths are
\begin{equation}
|M_F|^2={1\over 2J_i+1}|\langle J_f||
\sum_{i=1}^A\tau_+(i)||J_i\rangle|^2,
\end{equation}
and
\begin{equation}
|M_{\rm GT}|^2={1\over 2J_i+1}
|\langle J_f||\sum_{i=1}^A\sigma(i)\tau_+(i)||J_i\rangle|^2,
\end{equation}
respectively.
To evaluate Eq. (2) we must specify the distribution of these
transition probabilities over the 
final states of the daughter nucleus.
  
In the limit of good isospin the Fermi strength $|M_F|^2 = N-Z$
is carried entirely by the IAS in the daughter nucleus.
The excitation energy of the IAS, 
relative to the parent ground state, can be estimated
quite accurately from the Coulomb energy difference \cite{fuller}
\begin{equation}
E_{\rm IAS}\approx {1.728Z\over 1.12A^{1/3}+0.78}-1.293\ {\rm MeV}.
\end{equation}
The total GT strength is also simple due to the fact that the 
nuclei of interest have large neutron excesses, effectively 
eliminating all strength in the $(\bar\nu_e,e^+)$ channel.
It follows that the strength in the $(\nu_e,e^-)$ channel
is given by the Ikeda sum rule, 
$\sum_f |M_{\rm GT}|^2\approx3(N-Z)$.
However, the distribution of this GT strength is not determined
by general arguments and thus must be either calculated or
measured.  Studies using forward-angle $(p,n)$ scattering
with stable targets have shown that most of the strength is
concentrated in a broad resonance whose center is in the vicinity
of the IAS.  This motivated us to approximate 
the GT-strength distributions
for the nuclei of present interest as 
Gaussians centered at $E_{\rm GT}$
and with a full width at half maximum $\Gamma = 2(\ln2)^{1/2}\Delta$,
\begin{equation}
|M_{\rm GT}(E)|^2 \sim S\exp[-(E-E_{\rm GT})^2/\Delta^2].
\end{equation}
That is, we represent the distribution of $|M_{\rm GT}|^2$ summed
over all final states --- both discrete bound states and continuum
states --- by the continuous function $|M_{\rm GT}(E)|^2$ of Eq. (8).  
The normalization constant $S$ is fixed by the condition that Eq. (8),
integrated over all excitation energies $E \geq E_{\rm gs}$,
where $E_{\rm gs}$ is the ground state energy of the daughter nucleus,
gives the
sum rule result $3(N-Z)$.  We have taken the Gaussian centroids
$E_{\rm GT}$ from the analytic fit to 
$\delta = E_{\rm GT}-E_{\rm IAS}$
in Ref. \cite{naka}, where GT-strength 
distributions from forward-angle
$(p,n)$ measurements were studied.  While the measurements are 
confined to nuclei near stability, the data show that $\delta$
is linearly correlated with $N-Z$, leading to the prediction that
$\delta \sim 0$ for the neutron-rich nuclei of present interest.
We took $\Gamma \sim 5$ MeV
which is also a value typical of the 
$(p,n)$-measured GT-strength profiles.

Following the prescription outlined above,
we have calculated the rates for 
$\nu_e$ captures on the waiting-point nuclei
with $N=50$, 82, and 126 in the $r$-process, using an
average $\nu_e$ energy of 
$\langle E_{\nu_e}\rangle\approx 11$ MeV. The rates for $L_{\nu_e}=
10^{51}\ {\rm erg\ s}^{-1}$ at a radius 
$r=100$ km are given in Table I.
These rates can be easily scaled for different $L_{\nu_e}$
and $r$ using Eq. (1).

Such $\nu_e$ reactions typically excite the daughter nucleus to
states with excitation energies of $\sim 20$ MeV, 
which is well beyond particle breakup threshold.
Because these nuclei are very neutron-rich, 
they de-excite by emitting several neutrons, a process
we have simulated using a statistical neutron evaporation code
\cite{gavron}.  This code estimates the average number of neutrons
emitted by these nuclei, $\langle n \rangle$, as well as the
probabilities for emitting any specific number of neutrons (e.g.,
$P_{n=2}$), quantities that will be important in our
postprocessing calculations.  Nuclear masses, which are generally not
known experimentally, have been taken from the compilation of
M\"{o}ller {\it et al.} \cite{moller1}.
As expected from the average neutron separation energies at
the $N=50$, 82, and 126 shell closures, we find that
there are about 2--5 neutrons 
emitted after each $\nu_e$ capture on the waiting-point nuclei
(see Table I).

Neutral-current neutrino-nucleus interactions in supernovae
have been studied previously in 
Refs. \cite{haxton,kolbe1}.  As in the
charged-current case, there is an allowed contribution.  The
operator analogous to the Fermi operator contributes only to elastic
neutrino scattering, and hence is of no interest.  Thus inelastic
allowed transitions are governed by the neutral-current 
GT transition probability
\begin{equation}
|M_{\rm GT}^{\rm NC}|^2 ={1 \over 2J_i+1}
|\langle J_f||\sum_{i=1}^A\sigma(i){\tau_3(i)\over 2}||J_i\rangle|^2.
\end{equation}
We use the same renormalized $g_A^{\rm eff}$ as in the
charged-current case for the neutral-current GT transitions.
In the calculations described below we found that most of this
strength was concentrated below or very close to 
the neutron emission threshold, resulting in 
a relatively modest contribution to the neutron spallation 
channels of present interest.

Because $\nu_{\mu(\tau)}$ and $\bar\nu_{\mu(\tau)}$
have a higher mean energy,
odd-parity transitions generated by first-forbidden operators ---
those proportional either to the momentum transfer 
or to nucleon velocities --- must be considered.  
In Ref. \cite{haxton}
the first-forbidden contribution to neutral-current neutrino
scattering was calculated in a generalized Goldhaber-Teller model
satisfying the Thomas-Reiche-Kuhn (TRK) sum rule for E1 transitions
as well as its generalization for first-forbidden multipoles of
the axial current \cite{donnelly}.  
The strengths are carried by doorway states
placed in the center of the giant resonance region.  The TRK
sum rule predicts a response proportional 
to $NZ/A=(A/4)\{1-[(N-Z)/A]^2\}$. Even for the neutron-rich
nuclei in the $r$-process, $NZ/A\approx A/4$ 
is good within $\sim 10$\%. 
Therefore, if the first forbidden contribution
proves important, one expects the total heavy-flavor neutral-current
cross section to scale as
\begin{equation}
\langle\sigma_\nu^{\rm NC}\rangle\approx \sigma_0\times A,
\end{equation}
where $\sigma_0\approx 10^{-42}\ {\rm cm}^2$ 
for $\langle E_\nu\rangle
\approx 25$ MeV when averaged over, for example, 
$\nu_\mu$ and $\bar\nu_\mu$.  This was the case for the calculations 
done in Ref. \cite{haxton}, and it also was found to hold in our 
estimates of the cross section above neutron emission threshold
for nuclei of interest to the $r$-process.

We have tried to assess the accuracy of our various cross section
estimates by performing CRPA and shell-model ``benchmark" 
calculations.  The shell-model study was done for the waiting-point
nucleus $^{78}$Ni and its daughter $^{78}$Cu in the
$(\nu_e,e^-)$ reaction.
We employed a modified Kuo g-matrix (originally intended for the
$^{56}$Ni region, but rescaled by 0.8 to account for the larger
mass numbers of interest here), supplemented by the Sussex
potential matrix elements in those cases involving the
$1g_{7/2}$ shell.  The charged-current GT response was evaluated
from the naive $^{78}$Ni closed-shell ground state to a complete
set of $^{78}$Cu final states including, for example, those with 
a hole in the $1f_{7/2}$ shell or a particle in the $1g_{7/2}$
shell.  While this basis is somewhat simple, it has the virtue
that the Ikeda sum rule is exhausted by the $(\nu_e,e^-)$ 
channel; thus this calculation is compatible with the assumptions
made in our schematic treatment above.  The distribution of the
transition probability $|M_{\rm GT}|^2$ was evaluated in this
shell-model basis by a method of moments, 
rather than by a state-by-state
summation.  A least-squares best fit to the results 
using the function in Eq. (8)
was then made in order to determine shell-model values for the
parameters $E_{\rm GT}$ and $\Gamma$.  

The results are reasonably satisfying.  The shell model yielded
$\delta = E_{\rm GT}-E_{\rm IAS} = -3.36$ MeV, 
which can be compared with the
result of $-1.15$ MeV one obtains by 
extrapolating the fit of Ref. \cite{naka}.
Thus the prediction of this fit that the centroid
of the GT-strength distribution should 
fall below the IAS for the neutron-rich
nuclei of present interest is confirmed by the shell-model 
calculation.  While there is a difference of 
$\sim2$ MeV in the precise
location of the centroid, this is not very significant for our
neutrino cross sections: a shift of 2 MeV either way in the
centroid changes the cross sections by less than $\sim 50$\%.

The shell-model prediction for $\Gamma$, 11.7 MeV, is not in good
agreement with the assumed value of 5 MeV, which we argued was 
typical of fits to GT-strength profiles deduced from $(p,n)$ forward
scattering.  Because the neutrino cross sections are quite
insensitive to $\Gamma$, the origin of this discrepancy is a 
somewhat academic issue.  Nevertheless, these results motivated
us to perform analogous calculations for $^{64}$Ni, a stable
target for which the $(p,n)$-deduced 
GT-strength distribution is known \cite{rapa}.  Again
the shell-model prediction of $\delta\sim 5$ MeV was in good
agreement with the prediction of 3.4 MeV 
by the fit of Ref. \cite{naka},
as well as with the data. 
Yet the shell model yielded a less distinct resonance
than found experimentally, predicting concentrations of 
strength at both 7 and 17 MeV.  This suggested to us that our 
somewhat restricted shell-model basis might not include enough
of the correlations responsible for the GT resonance shape.
(This conclusion has been confirmed by recent shell model
Monte Carlo calculations performed in the full $2p1f$ shell which
can describe the measured GT-strength distributions
in Fe and Ni nuclei
reasonably well \cite{radha}.) Therefore
we concluded that the value of $\Gamma$ 
from $(p,n)$ systematics, 5 MeV, is
likely the more reliable choice.

Somewhat more sophisticated shell-model calculations were performed
for the allowed neutral-current cross section.  The $^{78}$Ni 
ground state was calculated in the $1f2p1g_{9/2}$ model space,
allowing all configurations with 0, 1, or 2 holes in the $1f_{7/2}$
shell.  The distribution of 
$|M_{\rm GT}^{\rm NC}|^2$ was again evaluated
by a method of moments, including configurations with 3 holes in the
$1f_{7/2}$ shell and one particle in the $1g_{7/2}$ shell.  
That is, a complete basis
was used for the final states.
The resulting sum rule strength for 
$|M_{\rm GT}^{\rm NC}|^2$ was 6.35,
yielding a quite large allowed cross section.  But the strength
was concentrated very near the ground state: 66.7\% was within 3 MeV,
80\% within 4 MeV, and 90\% within 6 MeV, which the mass formula
of Ref. \cite{moller1} indicates is the neutron separation energy.
Thus almost all of the strength is carried by bound states;
the allowed contribution to neutron spallation, 
the process of present
interest, is quite small, comparable to the forbidden contribution
discussed below. 
This supports the assumptions that led to Eq. (10).

We were able to extend the tests of neutral-current cross sections
to representative nuclei in each
of the three $r$-process abundance peaks by performing
CRPA calculations \cite{kolbe2}.  A Landau-Migdal
interaction was used and all multipoles of both parities through
$J=2$ were retained, thereby accounting for all 
allowed, first-forbidden, and second-forbidden operators.  
Thus the CRPA
calculations provide a check on the simpler Goldhaber-Teller
treatment of first-forbidden neutrino scattering 
\cite{haxton,donnelly}
and a cross-check on the shell-model result for the fraction
of allowed strength in the continuum.  The results for three
representative nuclei are listed in Table II.  They confirm the
simple scaling estimate in Eq. (10) to within $\sim40$\%.
Within this accuracy, the results are independent
of some of the existing neutrino spectrum uncertainties, such
as whether $\eta_\nu\sim 3$ is more appropriate than $\eta_\nu\sim0$.
The average number of neutrons emitted,
$\langle n\rangle$, and the probabilities for multiple neutron
emission, which are also listed in Table II, were obtained 
by folding the neutrino-induced excitation
spectrum calculated in the CRPA with the neutron-evaporation
spectrum determined from the statistical model \cite{gavron}.
We find that GT transitions
contribute about 40\% to the continuum cross sections, in agreement
with the shell-model result for $^{78}$Ni. Their contribution
to $\langle n\rangle$ is, however, less than 30\%
due to the lower excitation energies 
characterizing the allowed transitions. 
  
\section{Neutrino-nucleus interactions during the $r$-process}

The dynamic phase of the $r$-process is thought to occur between 
temperatures of $\sim3\times10^9$ and $\sim10^9$ K, during which time
$(n,\gamma)\rightleftharpoons(\gamma,n)$ equilibrium is maintained.
As photodisintegration reactions are typically orders of magnitude
faster than competing neutral-current neutrino spallation 
reactions, it is clear that inelastic neutrino scattering has 
no effect in the dynamic phase.  In contrast, charged-current 
neutrino reactions affect the charge flow and thus compete
with $\beta$-decay, particularly near the waiting points where
$\beta$-decay rates are anomalously small.  These charged-current
effects will be discussed in this section.  Below $\sim10^9$ K the 
material freezes out from
$(n,\gamma)\rightleftharpoons(\gamma,n)$ equilibrium,
fixing the distribution of the $r$-process progenitor 
nuclei, which decay back to the valley of $\beta$-stability to 
produce the abundances observed in nature.  After the freeze-out
neutrino interactions can affect the abundance distribution,
with both charged-current reactions and 
heavy-flavor neutral-current scattering
being important.
We will discuss several interesting consequences
of this neutrino postprocessing in Sec. IV.

Type-II supernovae have long been discussed as a possible site
of the $r$-process.  As mentioned in the introduction,
in the recently-developed $r$-process
model of Woosley {\it et al.} \cite{woosley} the synthesis occurs 
in the ``hot
bubble" expanding off the neutron star, with the freeze-out from
$(n,\gamma)\rightleftharpoons(\gamma,n)$ equilibrium occuring at
radii of 600--1000 km and at times $\sim10$ s after core
bounce.  However, despite the appeal of the hot bubble as an
$r$-process site, there are aspects of this model that are
unsatisfactory, such as overproduction of the isotopes 
$^{88}$Sr, $^{89}$Y, and $^{90}$Zr and the need for very
high entropies.  For this reason we will avoid the choice
of a specific $r$-process model, instead exploring a more
general scenario motivated by 
recent studies of nucleosynthesis in
a neutrino-driven wind blown off the neutron star \cite{qian}.  
Such a wind can be described as
an expanding bubble where
the material temperature decreases as $T\propto 1/r$
and the outflow velocity increases as $v\propto r$  
under the following assumptions: 
(1) the mass outflow rate is constant; (2) the expansion
of the radiation-dominated material in the outflow is
adiabatic; and (3) the
outflow is (barely) successful in ejecting
mass to infinity.
The time evolution for the radius of an expanding mass element 
in the outflow is given by
\begin{equation}
r(t) \propto \exp(t/\tau_{\rm dyn}),
\end{equation}
where $\tau_{\rm dyn}$ is a characteristic dynamic timescale for
the expansion.
We will denote the radius and neutrino luminosity for a mass element
at freeze-out, $T\sim 10^9$ K, by $r_{\rm FO}$ and $L_{\nu,\rm FO}$,
respectively.
We assume that the individual neutrino 
luminosities are the same
and generically denote $L_\nu$ as the luminosity for any 
neutrino species.  
The protoneutron-star neutrino luminosity is 
assumed to evolve with time
as $\exp(-t/\tau_\nu)$, with $\tau_\nu\sim3$ s.

In keeping with the notion that the discussion should be as 
general as possible, we will treat 
$\tau_{\rm dyn}$, $r_{\rm FO}$, and
$L_{\nu,\rm FO}$ as parameters relevant to the freeze-out of
a particular peak.  That is, their values for the $A\sim80$ 
($N=50$) peak could be
different from those for the $A\sim130$ or 195 ($N=82$ or 126)
peak.  This may be a prudent
generalization given the ongoing debate 
\cite{goriely} over whether $r$-process
abundances are consistent with a single production site. Furthermore,
even if there is only one $r$-process site,
because the neutron-to-seed ratio 
required to produce each peak is different,
individual peaks likely have to be made at
different times, and hence under different 
conditions in a single site. 

We first repeat the argument of Fuller and 
Meyer \cite{fuller}, who placed
a lower bound on the freeze-out radius by demanding that the
$r$-process must be in approximate steady-state local $\beta$-flow 
equilibrium at the time of freeze-out, 
a condition that Kratz {\it et al.} \cite{kratz}
deduced from the proportionality between the progenitor abundances 
along closed-neutron shells and the corresponding $\beta$-decay
lifetimes.  In local equilibrium, the product $\lambda(Z,N)Y(Z,N)$,
where $\lambda$ is the total charge-increasing rate and $Y$ the
abundance, should be independent of $(Z,N)$. 
The rate $\lambda$ includes
both $\beta$-decay and neutrino reactions, and if the latter 
are made too strong, the observed local equilibrium that holds
when only $\beta$-decay is considered is then destroyed.
Using the reaction rates in Table I,
the condition that local $\beta$-flow 
equilibrium holds to within 20\% at
freeze-out is especially restrictive at $N = 50$, yielding
\begin{equation}
\left({L_{\nu,\rm FO}\over 10^{51}\ {\rm erg}\ {\rm s}^{-1}}\right)
\left({100\ {\rm km}\over r_{\rm FO}}\right)^2 \lesssim 0.12\ 
{\rm for}\ A\sim80, 
\end{equation}
where this result corresponds to Fig. 2a of Ref. \cite{fuller}
[the comparison of $(Z,N) = (29,50)$ and $(30,50)$, which
generated the most stringent limit]
and depends on the $\beta$-decay rates in Table 4 of that reference.
Equation (12) is sufficient to guarantee 
that local $\beta$-flow equilibrium holds
at freeze-out provided that $\tau_{\rm dyn}$ 
is longer than but still comparable to typical
$\beta$-decay lifetimes, which are $\sim0.5$ s at $N=50$.
For the conditions in the $r$-process model of 
Woosley {\it et al.} \cite{woosley}, 
$L_{\nu,\rm FO} \sim 10^{51}$ erg s$^{-1}$ 
and $r_{\rm FO}\sim600$--1000 km,
Eq. (12) is easily satisfied.  But it is clear that
freeze-out radii of $\sim300$ km are also compatible with this
constraint.  If the calculation is 
repeated for the $N=82$ and 126 peaks
[corresponding to $(Z,N) = (45,82)$ and $(46,82)$, and to
$(67,126)$ and $(68,126)$, respectively, which we found
generated the most stringent limits],
the right-hand side of Eq. (12) becomes 0.83 and 0.37, respectively.
We have used the $\beta$-decay rates 
calculated by M\"oller {\it et al.}
\cite{moller2} for the $N=126$ nuclei. The numerical values for the
right-hand side of Eq. (12) may change somewhat if for example,
different $\nu_e$ energy spectra (cf. Ref. \cite{fuller})
are used to calculate the rates
in Table I. 

We now examine the effects of charged-current reactions prior to
freeze-out for the generic neutrino-driven wind $r$-process model
in order to assess whether neutrino interactions can speed up
the charge flow past waiting-point nuclei, given the constraint
imposed by Eq. (12).  The results in Table I show that 
charged-current reaction rates in the waiting-point regions
of $N =50$, 82, and 126 are reasonably constant, with 
$\bar{\lambda}_{\nu}^{\rm CC} = 
[(1/n)\sum^n_{i=1} {1/\lambda_i}]^{-1}\approx$
5.2, 5.7, and 8.5 s$^{-1}$, respectively, 
being average values at $r = 100$ km when
$L_\nu = 10^{51}$ erg s$^{-1}$.  
The number of transitions $\Delta Z_{\nu}$
induced by the neutrino irradiation is then
\begin{equation}
\Delta Z_{\nu} = \int_{t_i}^{t_f} \bar{\lambda}_{\nu}^{\rm CC}
\left[{L_\nu(t)\over 10^{51}\ {\rm erg}\ {\rm s}^{-1}}\right]
\left[{100\ {\rm km} \over r(t)}\right]^2 dt,
\end{equation}
where $t_i$ and $t_f$ are the times corresponding to the $r$-process
epoch between $3\times 10^9$ and $10^9$ K.
Under the generic wind scenario this can be evaluated to give
\begin{eqnarray}
\Delta Z_{\nu}& = &{9 \over 2} 
\bar{\lambda}_{\nu}^{\rm CC} \tau_{\rm dyn}
\left({L_{\nu,\rm FO} \over 10^{51}\ {\rm erg}\ {\rm s}^{-1}}\right)
\left({100\ {\rm km} \over r_{\rm FO}}\right)^2
{\exp[(\ln3)\tau_{\rm dyn}/\tau_{\nu}]-1/9 \over 
1 + \tau_{\rm dyn}/(2\tau_{\nu})}\nonumber\\
&\lesssim& 0.54 \bar{\lambda}_{\nu}^{\rm CC} \tau_{\rm dyn}
{\exp(1.1\tau_{\rm dyn}/\tau_{\nu})-1/9
\over 1 + \tau_{\rm dyn}/(2 \tau_{\nu})}
\ {\rm for}\ A\sim80,
\end{eqnarray}
where the last line follows from the $N=50$ 
freeze-out condition of Eq. (12).
Now this can be compared with 
the corresponding charge increase due to
$\beta$-decay,
\begin{equation}
\Delta Z_{\beta} = \bar{\lambda}_{\beta} (t_f - t_i) 
=(\ln3)\bar{\lambda}_\beta\tau_{\rm dyn}
\sim 1.1\bar{\lambda}_{\beta}\tau_{\rm dyn},
\end{equation}
where $\bar{\lambda}_{\beta} \sim 1.9\ {\rm s}^{-1}$ at $N=50$,
with the rates in Table 4 of Ref. \cite{fuller}.  Thus
\begin{equation}
{\Delta Z_{\nu} \over \Delta Z_{\beta}} 
\lesssim 0.49{\bar{\lambda}_{\nu}^{\rm CC}
\over \bar{\lambda}_{\beta}} 
{\exp(1.1\tau_{\rm dyn}/\tau_{\nu})-1/9 \over
1 + \tau_{\rm dyn}/(2 \tau_{\nu})}\ {\rm for}\ A\sim80.
\end{equation}
This result is quite interesting.  
It suggests that one can achieve an
$r$-process freeze-out pattern with characteristic local
$\beta$-flow equilibrium for the $N=50$ waiting-point nuclei
while still having
important --- even dominant --- neutrino contributions to the overall
$r$-process charge flow.  The above ratio is 
$\lesssim1.2$ under the assumption
$\tau_{\rm dyn}
\ll \tau_{\nu}$.  If $\tau_{\rm dyn}
\sim \tau_{\nu}$, this ratio increases by a factor of $\sim2$. 
If we repeat the calculation for $N=82$ (126), the 
numerical coefficient 0.49 on the
right-hand side of Eq. (16) becomes 3.4 (1.5) so that, 
with $\bar{\lambda}_{\beta} \sim4.3\ (16.0)\ {\rm s}^{-1}$
at $N=82$ (126),
$\Delta Z_{\nu}/\Delta Z_{\beta} \lesssim 4.0$ (0.71) when
$\tau_{\rm dyn} \ll \tau_{\nu}$.  Thus the charge flow can be totally
dominated by charged-current neutrino interactions in the $N=82$
peak without perturbing the local 
$\beta$-flow equilibrium at freeze-out
by more than 20\%.

The possible importance of this is clear.  The transit of the $N=50$
``bottle-neck'' requires a charge increase of
$\Delta Z \sim$ 5 and thus, if the flow is
carried only by $\beta$-decay, $\tau_{\rm dyn}$ must exceed 2.4 s,
according to Eq. (15).  This time is uncomfortably long for 
most neutrino-driven wind scenarios, 
where natural dynamic timescales are $\sim 0.1$--1 s \cite{qian}.
Of course, one may resort to a scenario where the nuclear flow
through the $N=50$ region is carried by $\alpha$-capture reactions
as in the $r$-process model of Woosley {\it et al.} \cite{woosley} to
accomodate such short timescales. However, even if the charge flow
is only carried by weak interactions,
the inclusion of charged-current neutrino reactions can
reduce the time required to clear the $N=50$ bottle-neck
by more than a factor of two without destroying
the $\beta$-flow equilibrium ``footprint" 
characterizing the freeze-out.
Similarly, the very large charge flow enhancements possible in
the $N=82$ peak allow, in principle, $\tau_{\rm dyn}$ 
as low as $\sim 0.2$ s. Finally, because most of the time
it takes to make the $N=126$ peak is spent at the $N=82$ and
possibly also the $N=50$ waiting-point nuclei on the $r$-process
path, the neutrino-induced charge flow enhancements for these nuclei 
corresponding to the limit on
the neutrino flux at the freeze-out of the $N=126$ peak, can
substantially reduce the required dynamic timescale for 
making this peak.
     
This is not the only attractive aspect of intense neutrino 
irradiation during the $r$-process: 
Fuller and Meyer \cite{fuller} pointed out
that charged-current reactions could help to correct the
overproduction of nuclei near $N=50$ in the $r$-process scenario
of Woosley {\it et al.} \cite{woosley} by 
increasing the electron fraction
mainly through $\nu_e$ capture on free neutrons. Furthermore,
they showed that some interesting light $p$-process nuclei could
be produced after including charged-current neutrino reactions on
nuclei in the reaction network.
Because the relative importance of $\nu_e$-induced
and $\beta$-decay-induced charge flow evolves during the $r$-process,
we may have reached a point where the explicit incorporation of
such effects into reaction network simulations of the $r$-process
is needed.  This may be essential to understanding the observed
pattern of abundances.
  
\section{Neutrino postprocessing effects}

As the temperature decreases to $\sim 10^9$ K, the material
freezes out from $(n,\gamma)
\rightleftharpoons (\gamma,n)$ equilibrium, leaving
a distribution of $r$-process progenitor nuclei which, after 
decay back to the valley of $\beta$-stability, produce the abundance
patterns seen in nature.

The charged-current reactions discussed in the previous section
can continue to influence the $r$-process abundance pattern during
postprocessing by altering the $(Z,N)$ path which the progenitors
follow as they $\beta$-decay.  The process of interest, neutron
emission following $(\nu_e,e^-)$ reactions, is superficially 
analogous to the $\beta$-delayed neutron emission process that 
is conventionally included in $r$-process calculations.  
However, the excitation energy of the daughter nucleus is 
significantly higher for neutrino reactions, leading to much
higher probabilities of multiple neutron emission, as can be seen
in Table I.  Thus the inclusion of charged-current reactions in
the postprocessing phase has the potential to push abundance
peaks to significantly smaller $A$.

Furthermore, since the postprocessing phase is defined by the
condition that $(n,\gamma) \rightleftharpoons (\gamma,n)$
equilibrium has broken down, 
the effects of neutral-current neutrino reactions 
are no longer competing with those of $(\gamma,n)$ reactions.
Nuclear excitation by $(\nu,\nu')$ reactions
above the particle breakup threshold may produce one or
more neutrons, again shifting abundances to lower $A$.  The
important species, due to their higher mean energies, are the
heavy-flavor neutrinos.
  
The optimal procedure for evaluating these neutrino irradiation
effects would be to incorporate them directly into the network
codes that follow the progenitors as they $\beta$-decay.  Our
procedures here are less sophisticated, though we would argue
that they are at least adequate qualitatively, given the other
uncertainties in $r$-process calculations.  We make three
simplifications.  First, as is apparent from Tables I and II,
neutrino rates and neutron spallation yields do not vary
excessively over an abundance peak.  For example, in the $N=50$
peak, $\lambda_\nu^{\rm CC}$ varies by about $\pm40$\%, while the
average number of neutrons emitted per neutrino event, 
$\langle n\rangle$, varies by about $\pm30$\%. (Variations between 
peaks are more significant: $\langle n \rangle$ for $N=126$ is
about twice that for $N=50$.)  Thus it is a reasonable approximation
to extract from Table I average values 
$\bar{\lambda}_\nu^{\rm CC}$, 
as we did in Sec. III, as well as averages
for the probabilities of emitting $n$ neutrons following a 
neutrino interaction, $\bar{P}_n^{\rm CC}$.  
Similarly, we take the values
in Table II as representative of the entire peaks 
near $N=50$, 82, and 126.  
Second, we make the additional simplification
that these mean progenitor rates and neutron emission probabilities
can be used throughout the postprocessing phase, even as $N-Z$
is evolving due to $\beta$-decay and 
charged-current neutrino reactions.  This is probably a quite good
assumption for neutral-current reactions because rates are tied
to the TRK sum rule and neutron emission probabilities to the
location of the giant resonances, both of which are only weakly
dependent on $N-Z$ for not too large changes in $A$.  
It is a more dangerous approximation in the
case of charged-current reactions because the available allowed
strength and $\langle n \rangle$ are strongly
correlated with $N-Z$.  Therefore, results for cases where
$\tau_{\rm dyn}$ is long compared with typical $\beta$-decay 
and/or $\nu_e$ capture rates
should be viewed with caution.  Third, we do not consider 
the subsequent processing of neutrons liberated in the spallation.
The neutron reabsorption process is quite different 
from the neutrino-induced multiple-neutron spallation
process, as only one neutron is captured at a time.
In addition, the
reabsorption should take place over the broad range
of nuclei with reasonable abundances and 
strong $(n,\gamma)$ cross sections
that reside below the abundance peaks, as well as on the
more plentiful nuclei with smaller but not unimportant
neutron capture 
cross sections that lie on the 
high-mass side of the abundance
peaks. This contrasts with the neutrino-induced neutron
spallation reactions, where dramatic effects occur only
for 3 or 5 nuclei concentrated in special ``windows.''
Therefore, while there may be consequences associated
with neutron recapture, we expect its net effect will 
be global and gentle, thus not undoing the neutrino
postprocessing effects described below.

With these approximations, the neutrino postprocessing effects 
can be evaluated without reference to the details of the $r$-process.
This is helpful in illustrating the kinds of effects that might
result from the neutrino irradiation.
The starting point is to calculate the mean number of neutrino events
$\bar{N}(n)$ producing exactly $n$ 
neutrons in the subsequent spallation
by integrating over all times after the freeze-out,
\begin{eqnarray}
\bar{N}(n)&=&\left(\bar{P}_n^{\rm NC}\bar{\lambda}_\nu^{NC}
+\bar{P}_n^{\rm CC}\bar{\lambda}_\nu^{\rm CC}\right) 
\left({L_{\nu,\rm FO} \over 10^{51}\ {\rm erg}\ {\rm s}^{-1}}\right)
\left({100\ {\rm km} \over r_{\rm FO}}\right)^2
\int_0^{\infty} \exp\left[-t\left({2\over\tau_{\rm dyn}} 
+ {1 \over \tau_{\nu}}\right)\right]dt\nonumber\\
&=&\left(\bar{P}_n^{\rm NC}\bar{\lambda}_\nu^{\rm NC}
+\bar{P}_n^{\rm CC}\bar{\lambda}_\nu^{\rm CC}\right) 
\left({L_{\nu,\rm FO} \over 10^{51}\ {\rm erg}\ {\rm s}^{-1}}\right)
\left({100\ {\rm km} \over r_{\rm FO}}\right)^2
{\tau_{\rm dyn}/2\over 1 + \tau_{\rm dyn}/(2\tau_{\nu})}.
\end{eqnarray}
Comparing the rates in Tables I and II, 
we see that $\bar{\lambda}_\nu^{\rm NC}$
is about 1.7 and 1.4 times $\bar{\lambda}_\nu^{\rm CC}$ 
for the $N=82$ and 126 peaks, respectively.
But the average number of neutrons emitted per 
charged-current interaction
is about twice that for neutral-current interactions in the
$N=126$ peak, a consequence of the 
large neutron excess and subsequent
high excitation energy of the daughter 
nucleus following $(\nu_e,e^-)$
interactions.  Thus charged-current 
and neutral-current interactions are
of comparable importance in driving neutron spallation near $N=126$.

Now in a neutrino-driven wind governed 
by $\tau_{\rm dyn}$ and $\tau_{\nu}$,
a given nucleus can interact one or more times, emitting several
neutrons.  We would like to calculate the net probability that,
at the end of postprocessing, a given progenitor nucleus has
lost $n$ neutrons.  The assumptions we have enumerated above make
this a straightforward exercise: 
rates and neutron emission probabilities
are not affected by the prior history of the target nucleus.  Thus
the distribution of events of each type --- e.g., the distribution
of neutrino events that produce exactly 
$n$ spallation neutrons --- is
governed by a Poisson distribution with parameter $\bar{N}(n)$.
The overall probability for emitting some number of neutrons
is given by counting up the number of ways this can be done
(e.g., two neutrons can be ejected by one neutrino interaction
that knocks out two, or by two interactions each of which knocks
out one), and folding the Poisson probabilities for each type
of event in the product.  The resulting distributions, which are 
not Poissonian, are tabulated in Tables III and IV for the $N=82$
and 126 peaks, respectively, and depend on 
the dimensionless parameter
\begin{equation}
{\cal{F}}=\left({L_{\nu,\rm FO} 
\over 10^{51}\ {\rm erg}\ {\rm s}^{-1}}\right)
\left({100\ {\rm km} \over r_{\rm FO}}\right)^2
\left({\tau_{\rm dyn}\over {\rm s}}\right)
{1\over 1+\tau_{\rm dyn}/(2\tau_{\nu})}.
\end{equation}
In the supernova $r$-process model of 
Woosley {\it et al.} \cite{woosley},
freeze-out occurs at radii of 600--1000 km over a 
dynamic timescale of $\tau_{\rm dyn} \sim
\tau_{\nu} \sim 3$ s. In this case, ${\cal{F}}$
would lie in the range 0.020--0.056.
Thus, for such long $\tau_{\rm dyn}$, appreciable neutrino
postprocessing can occur even at such large freeze-out radii,
as is apparent from Tables III and IV 
(and from the discussion below).

It is obvious from Eq. (18) that postprocessing effects depend
sensitively on $\tau_{\rm dyn}$, especially for 
$\tau_{\rm dyn}<\tau_\nu$.  
This was not the case for the
neutrino effects that occur prior to the freeze-out: the condition
of local $\beta$-flow equilibrium at 
freeze-out [Eq. (12)] constrains only
the neutrino luminosity and radius, or equivalently the neutrino flux
at freeze-out.  Likewise the fractional
increase in $r$-process charge flow due to neutrinos
[Eq. (16)] has a significant dependence on $\tau_{\rm dyn}$ only
for long dynamic timescales $\tau_{\rm dyn}\sim \tau_{\nu}$.

The proper use of the results in Tables III and IV is a 
nontrivial issue.  Neither our prejudices about $\tau_{\rm dyn}$ nor
the freeze-out constraints similar to
Eq. (12) significantly constrain the
neutrino fluence parameter $\cal{F}$ in Eq. (18).  
For a large fluence, a naive perturbative folding
of a calculated $r$-process abundance 
distribution with these spallation
probabilities could be misleading.  Worse, the calculated initial
distribution would have likely been ``tuned" to reproduce 
observation, fitting, for example, the abundance
peak at $N=82$.  But tuning prior to postprocessing is clearly 
inappropriate: one should strive to produce a best fit only 
after the final postprocessed distribution is achieved.

There is an alternative strategy that is appealing in its 
simplicity and avoids the need for a ``base-line"
unpostprocessed distribution from theory: begin with the
observed $r$-process abundance distribution and, 
for a given neutrino fluence,
invert this distribution to derive the yields that must have
existed prior to postprocessing.  This distribution would be
the one conventional theory should match, if indeed the neutrino
postprocessing effects are as described here.
Part of the appeal of this procedure is that the final $r$-process
abundance distribution is rather tightly 
constrained by observation and the neutrino
physics is relatively simple, 
governed by a single parameter $\cal{F}$ [Eq. (18)]
in the wind scenario.  Thus we can derive the unpostprocessed
distribution with some confidence. Note however, some caution
must be exercised when one compares the unpostprocessed
distribution derived this way with the 
progenitor abundance pattern at freeze-out calculated in
$r$-process models.
This is because effects of $\beta$-delayed neutron emission 
on the freeze-out pattern are hard to deconvolve in general,
although neutrino
postprocessing commutes with $\beta$-delayed neutron emission 
under the assumption of target-independent 
neutrino-induced neutron spallation.

In the region of a mass peak the inversion is relatively simple
to carry out iteratively.  As an initial guess for 
the unpostprocessed distribution,
we take the solar $r$-process abundances.
The postprocessing is calculated, and the deviations between the
resulting distribution and observed abundances are then used
to guess a new unpostprocessed distribution.  The procedure
is then iterated until the unpostprocessed distribution
converges, i.e., yields a postprocessed distribution in agreement
with observation. Depending on the neutrino fluence,
the final abundance of a particular
nucleus is affected by only a specific 
number of nuclei with higher masses. Therefore,
one can use an alternative inversion procedure by only
considering possible postprocessing contributions from 
nuclei below a cut-off high mass nucleus
sufficiently far away from the peak, 
say 10 mass units above the
peak nucleus. Starting from
the cut-off nucleus, one can calculate 
the unpostprocessed abundances of nuclei 
with successively lower masses.
These two procedures
yield the same results except near the cut-off
high mass nucleus. In other words, our postprocessing
results are insensitive to the choice of the cut-off,
so long as it is sufficiently far away from the peak.

\subsection{The N=82 peak}
  
The first result one can get from such an analysis is an upper
bound on the neutrino fluence parameter $\cal{F}$ of Eq. (18).  
The region of greatest sensitivity to the postprocessing are
those isotopes of low abundance lying just below the $N=82$ peak:
the inversion shows that the region $A=124$--126 is particularly
sensitive to the neutrino irradiation.  The
requirement that these isotopes not be overproduced provides a
stringent constraint on the neutrino fluence: if the parameter
$\cal{F}$ in Eq. (18) is made too large, 
the inversion gives unphysical (negative)
unpostprocessed abundances for these nuclei.

The deduced limit on the parameter of Eq. (18)
\begin{equation}
{\cal{F}}=\left({L_{\nu,\rm FO}
\over 10^{51}\ {\rm erg}\ {\rm s}^{-1}}\right)
\left({100\ {\rm km} \over r_{\rm FO}}\right)^2
\left({\tau_{\rm dyn}\over {\rm s}}\right)
{1\over 1 + \tau_{\rm dyn}/(2\tau_{\nu})}
\lesssim  0.09\ {\rm for}\ A\sim130
\end{equation}
is quite stringent.  For such a fluence, the neutrino postprocessing
contributions to the abundances of $^{124}$Te, $^{125}$Te, and
$^{126}$Te are 0.24, 0.45, and 0.65, respectively, which can be
compared with ranges deduced from observation, 0.215 $\pm$ 0.020,
0.269 $\pm$ 0.042, and 0.518 $\pm$ 0.126 \cite{kapp1}.  
Thus this fluence
is sufficient to overproduce all three isotopes, with the
$^{125}$Te discrepancy being particularly severe ($4\sigma$).

This limit, when combined with the freeze-out constraints we derived
following Fuller and Meyer \cite{fuller}, 
defines an allowed region of neutrino fluxes at freeze-out
and dynamic timescales, as shown in Fig. 1(a). In this figure, the 
horizontal solid line corresponds to the upper limit on the neutrino
flux at freeze-out similar to Eq. (12), 
but for the $A\sim130\ (N=82)$
peak. The diagonal solid line corresponds to the upper bound on the
neutrino fluence after freeze-out in Eq. (19) for the same peak.
The region bounded by these two lines gives the allowed conditions
at the freeze-out of the $A\sim 130$ peak. 

As the neutrino postprocessing calculations point to the
$A=124$--126 region as being most sensitive to such effects, it is
interesting to examine this region more carefully.  Using a 
fluence parameter of ${\cal{F}}=0.062$, which is compatible with
Eq. (19), one finds abundances of 0.18, 0.35, and 0.50 for
$A = 124$, 125, and 126, respectively.  As the agreement 
with the observed abundances given in the paragraph above is quite
good, the comparison hints that this region might be one where
neutrino postprocessing effects dominate the production.
This would, of course, be an exciting result as any determination
of the postprocessing fluence would quantitatively constrain
the location and dynamic timescale for the $r$-process, 
through an equality analogous to Eq. (19).
In this connection, we note that standard 
(unpostprocessed) $r$-process calculations
often significantly underestimate abundances 
in a relatively broad region around
$A\sim120$ \cite{kratz}, so there is room for additional production.
(As large postprocessing effects are confined to the region 
$A = 124$--126, they are not a solution for all of the deficiencies
in the $A\sim120$ region.  Effects such as shell quenching, 
which would revise the mass formulas commonly used, also help
to reduce the discrepancies \cite{chen}.)  

We performed the inversion 
under the assumption that the nuclei in the $A=124$--126 window are
attributable entirely to the postprocessing, that is, for 
a fluence parameter [Eq. (18)] of ${\cal{F}}=0.062$.  The solid line
in Fig. 2 is the resulting unpostprocessed distribution.
The $r$-process abundance distribution
deduced from solar abundances is also shown in this figure
as filled circles. The dashed line essentially
going through all the filled
circles is the
abundance pattern that results from 
the neutrino postprocessing of the
solid curve. To highlight the neutrino-induced 
synthesis of the nuclei in
the $A=124$--126 window, we blow up this region in the inset of
Fig. 2, and plot the error bars on the observed solar $r$-process
abundances. As we can see, all three nuclei are produced within
$\sim1\sigma$ of the observed abundances for a 
single neutrino fluence.

There are a few additional features clear from Fig. 2:

(1) The fluence parameter of ${\cal{F}}=0.062$ 
corresponds to an average neutron emission
number of $\langle n\rangle=1.05$.  
Thus one might expect to see a shift in the
peak of the distribution by this much.  This does not occur
because most nuclei (62\%) do not interact with the neutrino
flux: an $\langle n \rangle$ of 1.05 is achieved largely through
the emission of two or three neutrons by 
$\sim20$\% of the nuclei.

(2) Therefore, the signature of 
neutrino postprocessing is not a shift
in the peak, but rather a distortion of the shape of the peak.
Features tend to be more exaggerated before postprocessing:
The $A\sim130$ peak is higher before postprocessing, its edge on the
low-mass side is steeper, and the valley at $A=124$--126 is deeper.
Thus the net effect of the postprocessing on the lower two-thirds
of the abundance peak is not unlike that of pressing on a
steep pile of sand.

(3) In addition to spreading the abundance peak, the postprocessing
has a modest smoothing effect.  If one calculates the 
average magnitude of the ratio of the difference 
between neighboring peaks and
valleys to half their sum, it is 1.00 before postprocessing and 0.61
after.  These averages were evaluated in the region between masses
114 and 136.

\subsection{The N=126 peak}

The analogous inversion in the region of the $N=126$ abundance peak
again revealed a region on the low-mass tail of the peak where
postprocessing contributions are especially pronounced.  
This region spans the mass numbers $A=183$--187
and thus the nuclei $^{183}$W, $^{184}$W, $^{185}$Re, $^{186}$W,
and $^{187}$Re.  As in the $N=82$ peak we establish a conservative
upper bound on the neutrino fluence by finding the inversion
conditions under which all of these nuclei are overproduced 
by the postprocessing alone,
\begin{equation}
{\cal{F}}=\left({L_{\nu,\rm FO}
\over 10^{51}\ {\rm erg}\ {\rm s}^{-1}}\right)
\left({100\ {\rm km} \over r_{\rm FO}}\right)^2
\left({\tau_{\rm dyn}\over {\rm s}}\right)
{1\over 1 + \tau_{\rm dyn}/(2\tau_{\nu})}
\lesssim  0.06\ {\rm for} A\sim195.
\end{equation}
A fluence saturating this bound overproduces all five species,
with the deviations being $\gtrsim 3 \sigma$ in four cases
(and with the disagreement for $^{187}$Re being particularly
large, 0.067 compared with 0.0373 $\pm$ 0.0040 \cite{kapp2}).
The constraint in Eq. (20) can be combined with the freeze-out
bound of Sec. III again to severely limit the allowed 
neutrino flux at freeze-out
and the dynamic timescale. The results are given in Fig. 1(b).

The appearance of a well-defined region where neutrino postprocessing
effects are particularly pronounced suggests again that we test
the ansatz that these nuclei are entirely 
products of the postprocessing.
For a fluence parameter of ${\cal{F}}=0.03$ one obtains\newline
\indent
$A=183$~~~~0.0053~~~0.0067 $\pm$ 0.0016 \cite{kapp2}\newline
\indent
$A=184$~~~~0.0093~~~0.0135 $\pm$ 0.0035 \cite{kapp2}\newline
\indent
$A=185$~~~~0.0160~~~0.0127 $\pm$ 0.0024 \cite{kapp2}\newline
\indent
$A=186$~~~~0.0274~~~0.0281 $\pm$ 0.0024 \cite{kapp2}\newline
\indent
$A=187$~~~~0.0411~~~0.0373 $\pm$ 0.0040 \cite{kapp2}\newline
where the first number is the postprocessing result and the
second the abundance deduced from observation.
The correspondence is quite good and these results, especially 
when considered together with the $N=82$ results, are very suggestive.

We again stress that the regions where postprocessing effects are
most important, $A=124$--126 and 183--187, are clearly identified
by the inversion procedure, the input 
to which consists of the neutrino
cross sections and the associated multiple neutron 
emission probabilities
we have calculated and the $r$-process abundances
derived from observation.  The identification of 
these regions as sensitive 
to postprocessing does not, of course, 
require that the postprocessing
effects be large.  That will depend on where the $r$-process
occurs in the supernova --- or whether the supernova is even the
correct site.  But as we can do the inversion for any assumed 
neutrino fluence, the pattern of abundances in these regions can
either help to confirm or rule out the 
possibility of important neutrino-induced
synthesis.  We find that the observed abundance pattern
in both regions is characteristic of neutrino postprocessing.

Provided we have not been misled by an unfortunate conspiracy
of numbers, the conclusions 
would appear to be very important.  First, this suggests that a
core-collapse supernova (or at least some environment characterized
by a similarly intense neutrino fluence) is the site of the 
$r$-process.  Second, the required fluence parameters to produce the
$A=124$--126 and 183--187 isotopes can be calculated and are
${\cal{F}}=0.062$ and 0.03, respectively.  Thus we have been
able to place an important constraint on the $r$-process independent
of the many uncertainties that usually enter into network 
simulations.  As these fluences are modest, it appears 
either that the freeze-outs occur at large radii or that the 
dynamic timescales are short.  Most importantly, 
the derived postprocessing
fluence sharply constrains any model of 
the supernova $r$-process nucleosynthesis.  For
example, in the neutrino-driven wind model discussed, 
$\tau_{\rm dyn}$ is
now determined as a function of the neutrino flux at freeze-out,
or the freeze-out radius given the neutrino luminosity.
The third conclusion, which is more
uncertain, is that the factor of two difference in the $N=82$
and 126 postprocessing fluences suggests that
the $N=82$ peak freezes out at a smaller
radius than the $N=126$ peak 
in a dynamic $r$-process model such as the
neutrino-driven wind. In the wind models, 
large neutrino luminosities tend to
drive faster expansions of the outflow, and hence
correspond to shorter
dynamic timescales \cite{qian}.
As a result, the effects of $\tau_{\rm dyn}$ and 
$L_\nu$ on the fluence
parameter ${\cal{F}}$ nearly cancel. 
Therefore, a larger fluence implies a smaller
freeze-out radius. (The 
reason for being cautious with this conclusion is that the
determination of the $N=82$ fluence depends on only three isotopes,
so the possibility of an unfortunate neutron-capture 
mimicking of the $A=124$--126 postprocessed abundances is not
out of the question. In addition, a consistent set of neutrino
luminosity, dynamic timescale, 
and freeze-out radius corresponding to 
a specific fluence can only be obtained in a detailed model
of the neutrino-driven wind.)

In Fig. 3 we present the results of the inversion --- the $r$-process
abundance distribution before neutrino 
postprocessing that would reproduce
observation --- for the $N=126$ peak and
for a fluence parameter of ${\cal{F}}=0.03$.
The qualitative aspects of this distribution are quite similar 
to those found for the $N=82$ peak:  the features of the distribution 
before postprocessing are more distinct.  This fluence corresponds
to an average neutron emission number of $\langle n\rangle$ = 0.914.
But, as in the $N=82$ case, this is accomplished 
by a small fraction of the nuclei
emitting multiple neutrons after neutrino interactions: 74\% of
the nuclei experience no interactions.  Thus there is no shift
in the peak of the abundance distribution.
  
\section{Conclusions}

In this paper we have explored the consequences of neutrino 
irradiation during both the dynamic 
[$(n,\gamma)\rightleftharpoons(\gamma,n)$ equilibrium]
and postprocessing phases of
the $r$-process.  Our calculations are based on reasonable 
treatments of both charged-current and neutral-current responses
in the relevant nuclear mass regions, 
and include important contributions
from forbidden transitions in the case of $\nu_{\mu(\tau)}$ and
$\bar\nu_{\mu(\tau)}$ interactions.

Following Fuller and Meyer \cite{fuller} 
we used the $\nu_e$ capture rates
and the observation of approximate local $\beta$-flow equilibrium in
the mass peaks to constrain the radius (and neutrino luminosity) at
freeze-out.  We then showed that this constraint still allows 
important --- in fact, dominant in the case of the $N=82$ peak ---
neutrino contributions to the overall $r$-process charge flow.
This is an interesting result since the times required to 
cross the $N=50$ and 82 peaks are, in the absence of neutrino effects,
uncomfortably long.

We then studied the postprocessing phase.  Because the neutrino
effects are relatively well understood, we argue that the
unpostprocessed abundance distribution can be reasonably well 
determined from observation as a function of the neutrino
fluence.  The result is the identification of two regions,
$A=124$--126 and 183--187, that are particularly sensitive to
neutrino postprocessing.  Furthermore, the pattern of abundances 
in both regions corresponds closely to that from neutrino-induced
neutron spallation.  Thus there is strong evidence that these eight 
isotopes are mainly produced by neutrino postprocessing.

If this chain of argument is correct, the $r$-process must
take place in an intense neutrino fluence.  This supports the
growing prejudice for a site within a core-collapse supernova.
The allowable dynamics of such supernova models are now
sharply constrained by the deduced postprocessing fluence:
for a given freeze-out radius and neutrino luminosity, the dynamic
timescale is determined.  A comparison of the fluences for
the $N=82$ and 126 peaks also hints of a dynamic $r$-process
where the $N=82$ peak freezes out at a smaller radius.

Although the deduced fluences dominate the nucleosynthesis 
only in the special regions of $A = 124$--126 and 183--187,
their effects elsewhere are not insignificant.  The 
features of the unpostprocessed distributions are significantly
more pronounced than those of the final distributions.  Thus
if one is interested in supernova $r$-process sites with even
modest neutrino irradiation, it is unwise to tune network
simulations to reproduce final $r$-process abundance distributions 
unless the neutrino effects have been evaluated.

The present calculations involved several ``short cuts" that,
though reasonable, should be re-examined in future calculations.
We believe our results provide ample motivation for a full 
inclusion of neutrino effects in $r$-process networks.

\acknowledgments
 
We thank George Fuller, Brad Meyer, and 
Friedel Thielemann for helpful discussions.
Y.-Z. Qian and P. Vogel acknowledge the 
Institute for Nuclear Theory at University of Washington,
Seattle for its hospitality during the time part of
this work was done.
This work was supported by the Department of Energy under Grant
Nos. DE-FG06-96ER40561 and DE-FG03-88ER-40397, 
and by the National Science Foundation
under Grant Nos. PHY94-12818 and PHY94-20470.
Y.-Z. Qian was supported by the D. W. Morrisroe Fellowship
at Caltech.

\pagebreak

\pagebreak

\begin{figure}
\caption{Constraints imposed on the neutrino flux parameter
$L_\nu/r^2$ at freeze-out
and dynamic timescale $\tau_{\rm dyn}$ by the conditions
(horizontal solid lines) of local
$\beta$-flow equilibrium at the freeze-out of 
(a) the $A\sim130$ peak, and 
(b) the $A\sim195$ peak, and by the conditions 
(diagonal solid lines) that
neutrino postprocessing not overproduce nuclei in the regions of
(a) $A=124$--126, and (b) $A=183$--187. 
We have taken $\tau_{\nu}=3$ s.
Parameters lying on the dashed lines correspond to the fluences
determined by attributing the synthesis of nuclei in the
special mass regions to neutrino postprocessing (see text).}
\end{figure}

\begin{figure}
\caption{The unpostprocessed distribution 
(solid line) obtained from
the observed $r$-process abundances of Ref. [24] 
by unfolding the neutrino postprocessing effects.  We have
chosen a fluence parameter [Eq. (18)] of ${\cal{F}}=0.062$,
which provides a best fit to the observed abundances of
the nuclei with $A=124$--126
as highlighted in the inset.
The filled circles and error bars are data
taken from Ref. [24].
The dashed line essentially 
going through all the filled circles
corresponds to
the postprocessed distribution.}
\end{figure}

\begin{figure}
\caption{As in Fig. 2, but for the $N=126$ region.  
A neutrino fluence parameter [Eq. (18)]
of ${\cal{F}}=0.03$ 
has been used, as required
by the neutrino postprocessing fit to the $A=183$--187 region.
Solar $r$-process abundances are taken from Ref. [24] except
for the $A=182$--189 region, where the revised data of
Ref. [26] have been used.}
\end{figure}

\pagebreak

\begin{table}
\caption{Results for $\nu_e$ capture on the
waiting-point nuclei in the $r$-process.
The $\nu_e$ capture rates $\lambda_\nu^{\rm CC}$ are calculated 
with $\langle E_{\nu_e}\rangle\approx 11$ MeV, 
$\eta_{\nu_e}\approx3$, 
and $L_{\nu_e}=10^{51}\ {\rm erg\ s}^{-1}$ at
$r=100$ km, under the
assumption of a Gaussian GT-strength
distribution with a full width at half maximum
of $\Gamma\approx 5$ MeV.
The last seven columns give $\langle n\rangle$, 
the average number of neutrons emitted 
after a $\nu_e$ capture reaction,
and various probabilities for multiple
neutron emission.}
\begin{tabular}{ccccccccccc}
$Z$&$N$&$A$&$\lambda^{\rm CC}$ (s$^{-1}$)&
$\langle n\rangle$&$P_{n=1}$&$P_{n=2}$&$P_{n=3}$&
$P_{n=4}$&$P_{n=5}$&$P_{n=6}$\\\hline
26&50&76&8.1&3.1&0&0.06&0.83&0.08&0.02&0\\
27&50&77&7.0&2.9&0&0.11&0.83&0.06&0&0\\
28&50&78&6.0&2.0&0&0.88&0.12&0&0&0\\
29&50&79&5.1&2.0&0&0.92&0.08&0&0&0\\
30&50&80&4.3&1.9&0.11&0.89&0&0&0&0\\
31&50&81&3.7&1.8&0.19&0.81&0&0&0&0\\
31&52&83&4.5&2.0&0.02&0.89&0.09&0&0&0\\
%&&&&&&\\
45&82&127&7.3&3.0&0&0.03&0.91&0.06&0&0\\
46&82&128&6.5&2.6&0&0.32&0.68&0&0&0\\
47&82&129&5.8&2.5&0&0.45&0.55&0&0&0\\
48&82&130&5.2&2.1&0&0.83&0.17&0&0&0\\
49&82&131&4.6&2.0&0&0.90&0.10&0&0&0\\
49&84&133&5.3&3.0&0&0.04&0.95&0.01&0&0\\
%&&&&&&\\
65&126&191&10.1&5.1&0&0&0&0.10&0.75&0.15\\
66&126&192&9.3&4.6&0&0&0&0.42&0.58&0\\
67&126&193&8.5&4.6&0&0&0&0.38&0.62&0\\
68&126&194&7.8&4.0&0&0&0.08&0.83&0.09&0\\
69&126&195&7.2&4.0&0&0&0.06&0.89&0.05&0\\
\end{tabular}
\end{table}

\pagebreak

\begin{table}
\caption{Some illustrative results for 
neutral-current neutrino interactions
from CRPA calculations.  The cross sections 
(per heavy-flavor neutrino species)
above neutron emission threshold, 
$\langle\sigma_\nu^{\rm NC}\rangle$,
have been calculated with $\langle E_\nu\rangle
\approx 25$ MeV and $\eta_\nu\approx 0$.
The corresponding interaction rates $\lambda_\nu^{\rm NC}$ 
are summed over four heavy-flavor neutrino species,
and evaluated at $r=100$ km for a luminosity of 10$^{51}$
erg s$^{-1}$ per neutrino species.  The last seven columns give
$\langle n\rangle$, the average number of neutrons emitted after
a neutral-current neutrino interaction, and various 
probabilities for multiple
neutron emission.}
\begin{tabular}{cccccccccccc}
$Z$&$N$&$A$&
$\langle \sigma_\nu^{\rm NC}\rangle/A\ (10^{-42}\ {\rm cm}^2)$&
$\lambda_\nu^{\rm NC}$ (s$^{-1}$)&$\langle n\rangle$
&$P_{n=1}$&$P_{n=2}$&$P_{n=3}$&$P_{n=4}$&$P_{n=5}$&$P_{n=6}$\\\hline
28&50&78&0.56&3.5&2.0&0.45&0.26&0.16&0.11&0.02&0\\
48&82&130&0.94&9.7&2.0&0.37&0.40&0.14&0.08&0.01&0\\
68&126&194&0.75&11.6&2.0&0.41&0.37&0.10&0.08&0.03&0.01\\
\end{tabular}
\end{table}

\pagebreak

\begin{table}
\caption{Postprocessing neutron emission 
probabilities in the $A\sim130$
region. The fluence parameter ${\cal{F}}$ is defined in Eq. (18).
The second column gives $\langle n \rangle$, 
the average number of neutrons emitted by
a nucleus throughout the postprocessing, e.g., allowing for
the possibility of multiple neutrino interactions.
Both charged-current and neutral-current interactions are included.}
\begin{tabular}{ccccccccccccc}
${\cal{F}}$&$\langle n \rangle$&
$P_{n=0}$&$P_{n=1}$&$P_{n=2}$&$P_{n=3}$&$P_{n=4}$&$P_{n=5}$&$P_{n=6}$&
$P_{n=7}$&$P_{n=8}$&$P_{n=9}$&$P_{n \geq 10}$\\\hline
0.02&0.339&0.857&0.031&0.053&0.043&0.010&0.003&0.002&0.001&0&0&0\\
0.03&0.508&0.794&0.043&0.073&0.061&0.017&0.007&0.003&0.001&0&0&0\\
0.04&0.677&0.735&0.053&0.091&0.077&0.023&0.011&0.006&0.002&0.001&0&0\\
0.06&1.016&0.630&0.069&0.119&0.103&0.037&
0.021&0.012&0.005&0.002&0.001&0.001\\
0.08&1.355&0.540&0.078&0.137&0.123&
0.051&0.033&0.020&0.009&0.005&0.002&0.002\\
0.10&1.693&0.463&0.084&0.148&0.137&
0.063&0.045&0.029&0.015&0.008&0.004&0.004\\
0.12&2.032&0.397&0.086&0.154&0.146&
0.075&0.056&0.038&0.021&0.012&0.007&0.008\\
\end{tabular}
\end{table}

\pagebreak

\begin{table}
\caption{As in Table III, but for the $A\sim195$ region.}
\begin{tabular}{ccccccccccccc}
${\cal{F}}$&$\langle n \rangle$&
$P_{n=0}$&$P_{n=1}$&$P_{n=2}$&$P_{n=3}$&$P_{n=4}$&$P_{n=5}$&$P_{n=6}$&
$P_{n=7}$&$P_{n=8}$&$P_{n=9}$&$P_{n \geq 10}$\\\hline
0.02&0.609&0.818&0.039&0.036&0.013&
0.043&0.036&0.007&0.002&0.002&0.002&0.002\\
0.03&0.914&0.740&0.053&0.050&0.019&
0.059&0.051&0.012&0.005&0.004&0.004&0.003\\
0.04&1.218&0.669&0.064&0.061&0.024&
0.073&0.064&0.017&0.008&0.006&0.007&0.007\\
0.06&1.827&0.547&0.078&0.076&0.033&
0.092&0.084&0.029&0.016&0.013&0.013&0.019\\
0.08&2.436&0.448&0.086&0.085&0.040&
0.104&0.097&0.041&0.024&0.020&0.021&0.034\\
0.10&3.045&0.366&0.087&0.089&0.045&
0.110&0.106&0.051&0.033&0.028&0.029&0.056\\
0.12&3.654&0.300&0.086&0.090&0.048&
0.112&0.111&0.060&0.041&0.035&0.036&0.081\\
\end{tabular}
\end{table}

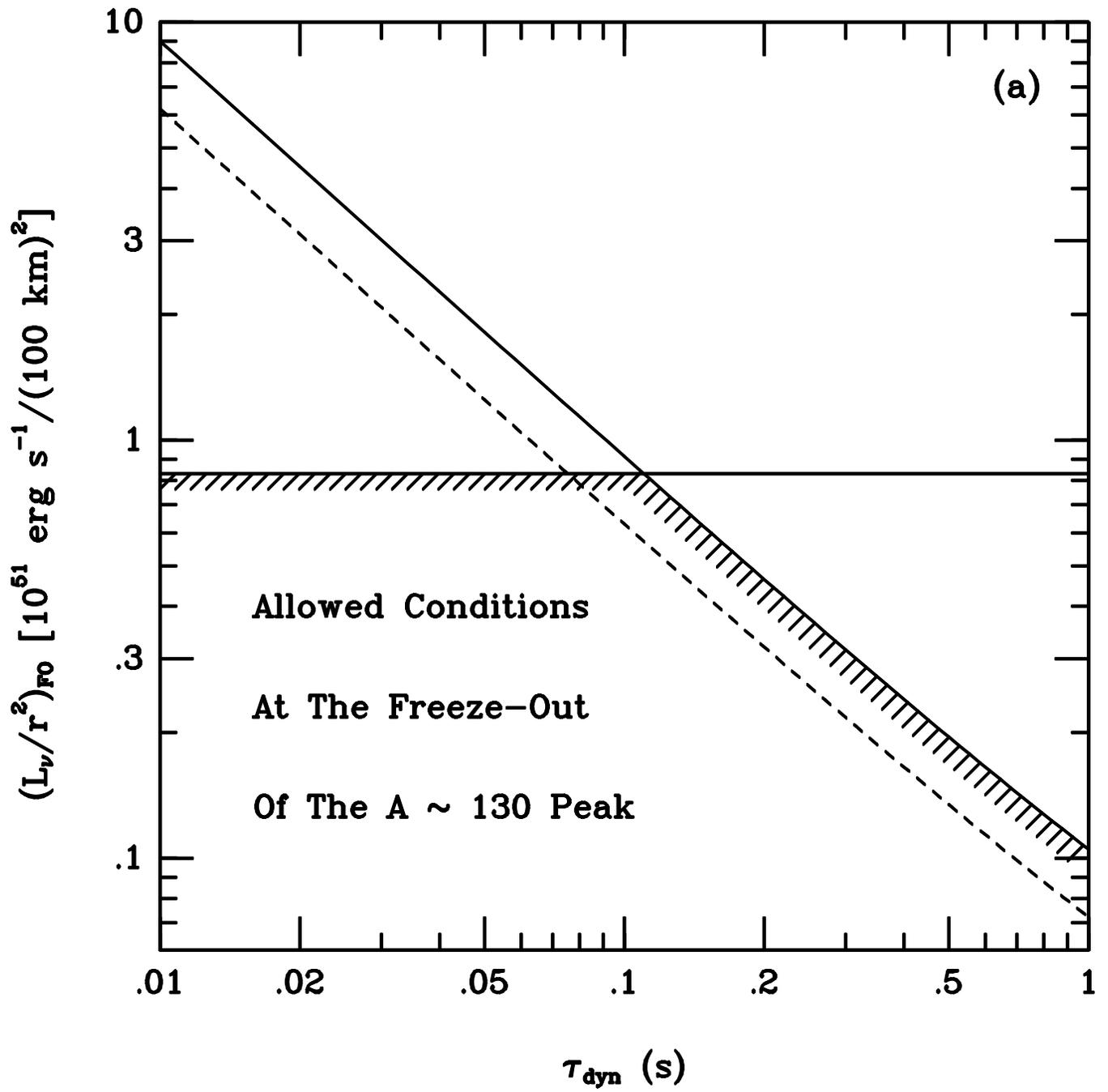
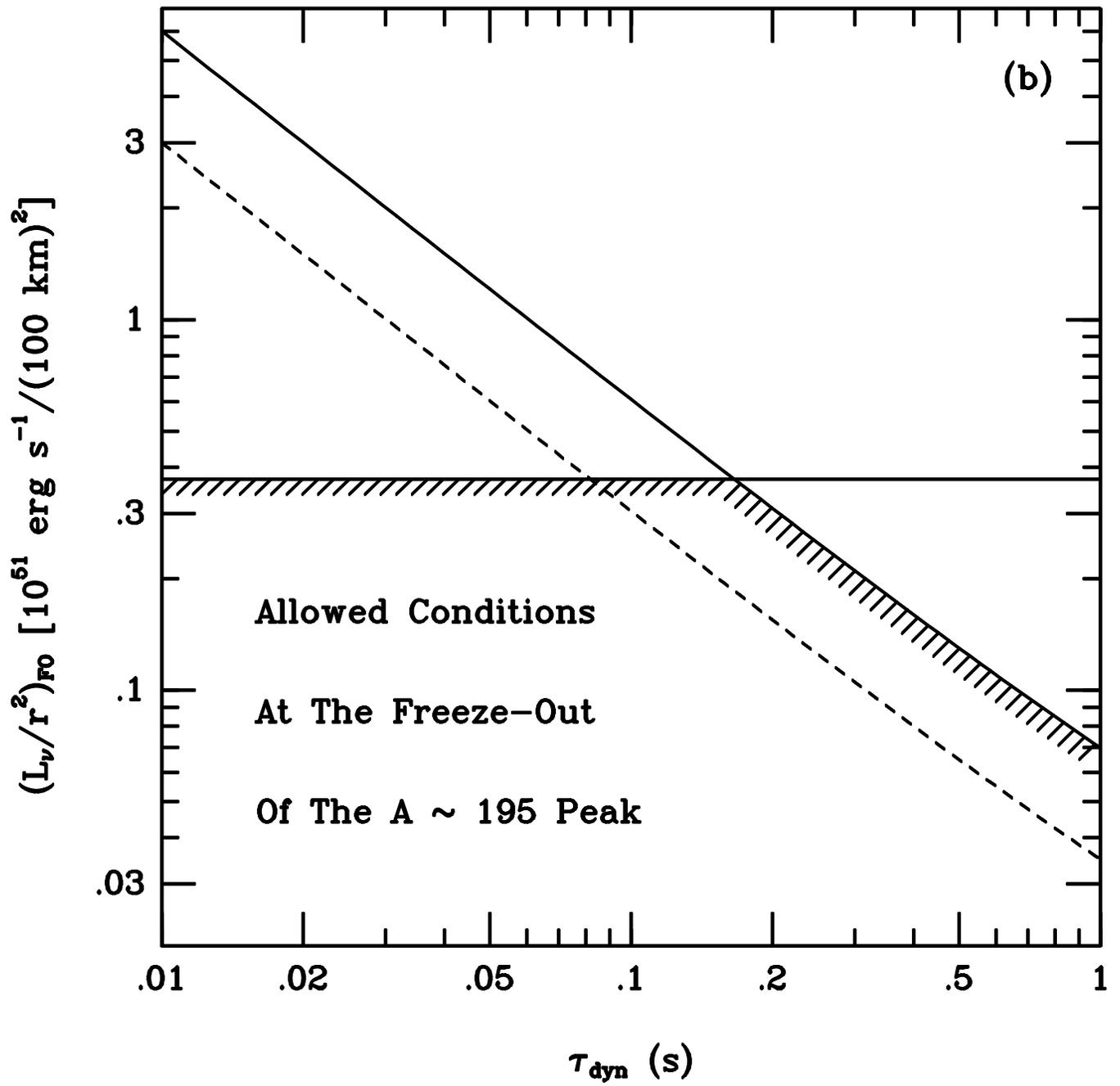
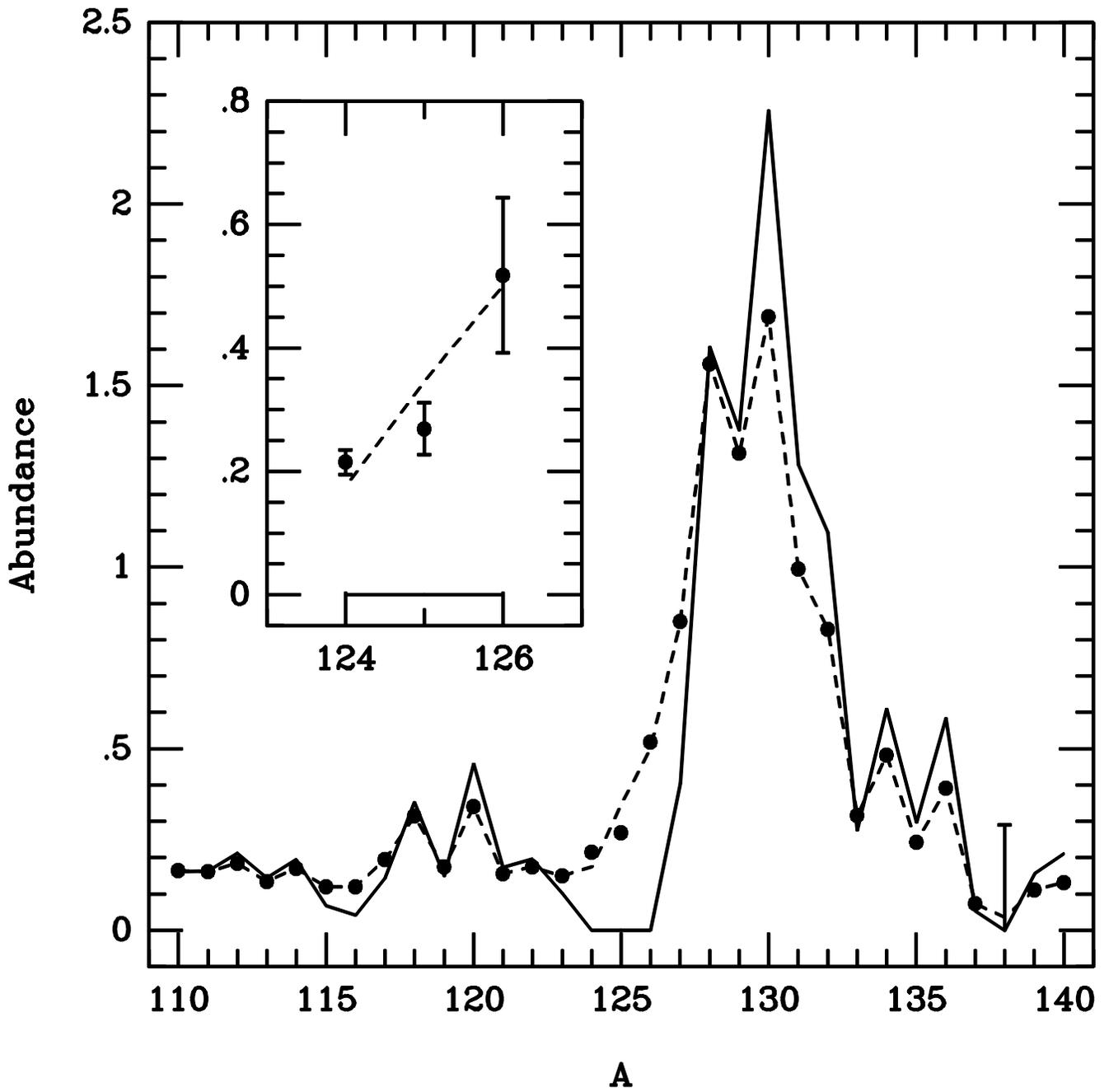
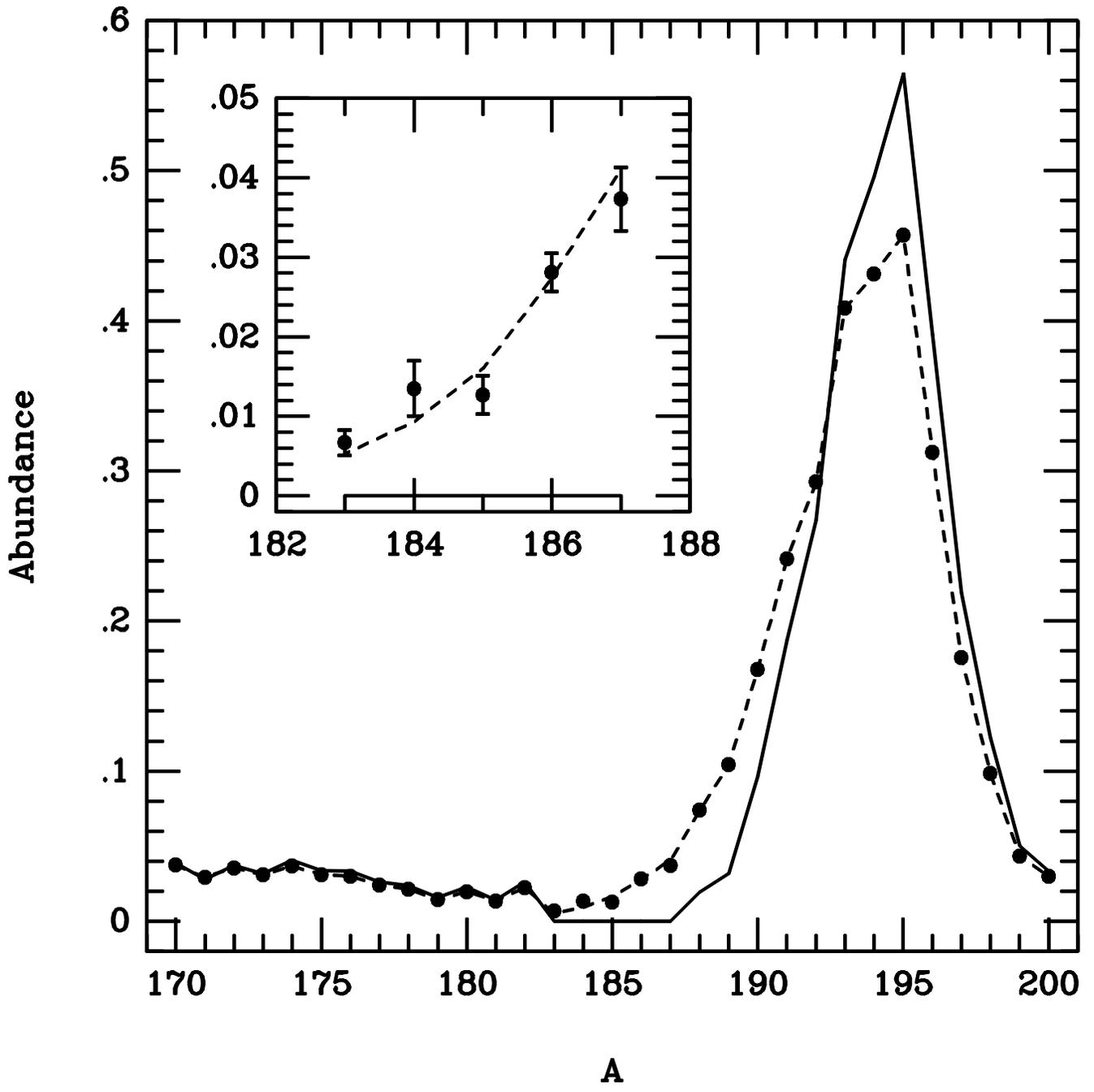
\end{document}